\documentstyle[12pt]{article}

\catcode`\@=11
\newbox\tempboxa
\newdimen\captionboxsubcount
\def\capsize#1{\captionboxsubcount=#1pt}
\newdimen\captionboxsub
\captionboxsub=\hsize \advance\captionboxsub by -\captionboxsubcount
\advance\captionboxsub by -\captionboxsubcount
\long\def\@makecaption#1#2{
 \setbox\@tempboxa\hbox{#1: #2}
 \ifdim \wd\@tempboxa >\captionboxsub
\rightskip=\captionboxsubcount \leftskip=\captionboxsubcount #1: #2
\else \hbox to\hsize{\hfil\box\@tempboxa\hfil}
 \fi}
\catcode`\@=12
\capsize{30}

\begin{document}

\begin{titlepage}
\begin{flushright}
OU-HET 253 \\
hep-th/9610211 \\
\end{flushright}

\begin{center} \LARGE
  On Instanton Calculations of \\
  ${\cal N}=2$ Supersymmetric Yang-Mills Theory
\end{center}
\bigskip

\begin{center} \Large
        Yuhsuke Yoshida
\footnote{e-mail address :
{\tt yoshida@fuji.phys.wani.osaka-u.ac.jp}} \\
\end{center}
\bigskip

\begin{center} \large \it
         Department of Physics \\
         Graduate School of Physics, Osaka University \\
         Machikaneyama 1-16, Toyonaka \\
         Osaka 560, JAPAN
\end{center}

\begin{center} \Large \bf
Abstract
\end{center}

\begin{quote}
Instanton calculations are demonstrated
from a viewpoint of twisted topological field theory.
Various properties become manifest such that
perturbative corrections are terminated at one-loop, and
norm cancellations occur between bosonic and fermionic
excitations in any instanton background.
We can easily observe that
for a suitable choice of Green functions
the infinite dimensional path integration
reduces to a finite dimensional integration
over a supersymmetric instanton moduli space.
\end{quote}


\end{titlepage}

\section{Introduction}
\label{intro}

In ${\cal N}=2$ supersymmetric quantum field theories
an simple way is given of how to determine
the low energy effective theory quite precisely.\cite{SW,SW2}
Holomorphy and symmetries severely constrain the possibilities
of quantum corrections.
It was also suggested that the prepotential of
low energy effective theory solely arises
from contributions of one-loop perturbations
and instantons.\cite{Seiberg}

An analysis indicated that there exists
one instanton contributions.\cite{Seiberg}
Using factorization property of gauge invariant operators,
this contribution is calculated
from a two-point Green function
$\langle{\rm tr}\phi^2(x){\rm tr}\phi^2(y)\rangle$
and the result is shown to be consistent with
its direct calculation of $\langle{\rm tr}\phi^2\rangle$
in the weak coupling limit.\cite{FiPo}
Here, $\phi$ is the Higgs field.
This indicates that Green functions
can be calculated in the weak coupling limit.
Upon this observation many calculations
of instanton contributions
have been done in the weak coupling
limit.\cite{DKM,FuTr,ItSa,SQCD}
Those results indicates that
non-perturbative contributions
seem to be saturated only by instantons.

In the present paper
${\cal N}=2$ supersymmetric Yang-Mills theory
is analyzed in terms of twisting and instanton calculations.
We easily show that calculations in the weak coupling limit
are guaranteed for a class of gauge invariant operators,
and non-perturbative contributions
are indeed saturated only by instantons.
Seen from a viewpoint
of twisted topological theory,\cite{Witten}
miraculous cancellations between
bosonic and fermionic excitations occur
for the class of Green functions.
Ultimately and undoubtedly,
an infinite dimensional path integration
reduces to a finite dimensional integration over
a supersymmetric instanton moduli space.
With these observations at hand,
we demonstrate a one-instanton calculation in terms of
the twisted topological field theory.
Twisting is a powerful tool to reveal various properties
of ${\cal N}=2$ supersymmetric instantons.

We work in a Coulomb branch
where a gauge symmetry breaking occurs.
In this type of theories instantons do not appear
in a simple way.
Some zero modes due to instantons are raised
by the Higgs mechanism.\cite{'t Hooft,Affleck,ADS}
In general an self-dual instanton solution
and the super-partner are specified
by a positive integer $k$
and have $8k$ degrees of freedom, respectively.
After the gauge symmetry breaking,
$8k-4$ of such zero modes are raised,
which we call quasi-zero modes.

This paper is organized as follows.
In section~\ref{SYM} we show basic ingredients of
the ${\cal N}=2$ supersymmetric Yang-Mills theory.
Especially, a viewpoint of topological field theory
is introduced.
Norm cancellations of massive modes are explained
in section~\ref{Q}.
In section~\ref{SIC} we demonstrate a instanton calculation
using the twisted topological field theory.
Summary and Discussion are found in the last section.

\section{${\cal N}=2$ Supersymmetric Yang-Mills Theory}
\label{SYM}

In this section we show basic ingredients of
the ${\cal N}=2$ supersymmetric Yang-Mills theory.
Especially, a viewpoint of topological field theory
is introduced.
This is utilized in supersymmetric instanton calculations,
and is powerful tool to reveal various properties.

\subsection{the Lagrangian}
\label{the L}

We consider supersymmetric $SU(2)$ Yang-Mills theory
whose Lagrangian respects the ${\cal N}=2$ supersymmetry
with vanishing central charge.
The Lagrangian has a global isospin symmetry $SU(2)_I$.
The supercharges
$Q_{\alpha}{}^i$ and $\overline Q_{\dot\alpha j}$
transform in the fundamental representation of $SU(2)_I$.
The Yang-Mills gauge field $A_m$ is embedded in the
${\cal N}=2$ chiral multiplet ${\cal A}$ consisting of
one ${\cal N}=1$ chiral multiplet
$\Phi = (\phi, \psi_\alpha)$ and
one ${\cal N}=1$ vector multiplet
$W_\alpha = (\lambda_\alpha, A_m)$.
The ${\cal N}=2$ chiral multiplet is arranged
as a diamond form
\begin{equation} \label{diamond}
\begin{array}{ccc}
 & A_m^a & \\
\lambda_\alpha^a & & \psi_\alpha^a\\
 & \phi^a & \\
\end{array}
\end{equation}
to exhibit the $SU(2)_I$ symmetry which acts on the rows.
The ${\cal N}=2$ supersymmetry determines
all the relations between the kinetic and interaction terms
up to a gauge coupling constant.
All the fields are introduced so that
the supersymmetry algebra does not contain
the gauge coupling constant.
We postpone introducing the gauge coupling constant
until carrying out path integrations.
In ${\cal N}=1$ superspace, the Lagrangian is
\begin{equation} \label{L}
{\cal L} =
\int d^4\theta\: \overline\Phi e^{-2V}\Phi e^{2V} ~+~
\frac{1}{4}\int d^2\theta\: W^\alpha W_\alpha ~+~
\frac{1}{4}\int d^2\overline\theta\:
   \overline W_\alpha \overline W^\alpha ~.
\end{equation}
We suppress gauge group indices,
for example $W^{a\alpha}W^a_\alpha = W^\alpha W_\alpha$,
otherwise stated.
The covariant derivative is defined
on adjoint representations
by $D_m\phi^a \equiv \partial_m\phi^a +
\epsilon^{abc}A_m^b\phi^c$.
Our notations will be found by putting
the gauge coupling constant as $g=-1$
in a standard book\cite{Wess-Bagger}.

Let us recall properties known from looking at
the supersymmetric theory (\ref{L}).

In terms of ${\cal N}=1$ superfield formalism,
only a supgroup of $SU(2)_I$, which is called $U(1)_J$,
is a manifest symmetry.
The $U(1)_J$ transformation acts on the $SU(2)_I$ doublet
in a diagonal form;
$\Psi\to \Psi(e^{-i\gamma}\theta)$,
$W_\alpha\to e^{i\gamma}W_\alpha(e^{-i\gamma}\theta)$.
The symmetries $SU(2)_I$ and $U(1)_J$ are non-anomalous ones
existing in the quantum level.

There is also an Abelian symmetry $U(1)_R$, which ensures
that the gaugino $\lambda_\alpha$
and the Higgsino $\psi_\alpha$
are bare massless, acting as
$\Psi\to e^{2i\gamma}\Psi(e^{-i\gamma}\theta)$,
$W_\alpha\to e^{i\gamma}W_\alpha(e^{-i\gamma}\theta)$.
The $U(1)_R$ symmetry is generally broken by an anomaly
down to a discrete symmetry ${\bf Z}_{4N_c}$.
A ${\bf Z}_2$ subgroup of the discrete R symmetry
also contains in the $U(1)_J$ symmetry.
Namely,
\begin{equation}
U(1)_R : \begin{array}{ccc}
 & A_m & \\
\omega\lambda_\alpha & & \omega\psi_\alpha\\
 & \omega^2\phi & \\
\end{array}
\quad = \quad
U(1)_J : \begin{array}{ccc}
 & A_m & \\
\omega\lambda_\alpha & & \omega^{-1}\psi_\alpha\\
 & \phi & \\
\end{array}
\end{equation}
if $\omega^2=1$.
In this case the global symmetry is
$SU(2)_R\times {\bf Z}_8/{\bf Z}_2$.

The classical potential of the theory (\ref{L}) is
\begin{equation}
V = \frac{1}{2}[\phi,\overline\phi]^2 ~.
\end{equation}
This potential has, so-called, $D$-flat directions in which
we have vanishing vacuum energy $V=0$.
The $D$-flatness implies that
the Higgs fields $\phi$ and $\overline\phi$
belong to a Cartan subalgebra of
the gauge symmetry $SU(2)_G$ developing
a vacuum expectation value
\begin{equation} \label{vev}
\langle\phi\rangle = \frac{1}{2}
\left(\begin{array}{cc} a & 0 \\ 0 & -a \end{array}\right) ~,~
\qquad a \in {\bf C^*} \equiv {\bf C}/\{0\} ~.
\end{equation}
This form is given after divided by a gauge transformation
continuously connecting to unity.
The Weyl group of $SU(2)_G$ acts by $a\to-a$,
so a convenient $SU(2)_G$ gauge invariant operator
parameterizing the space of vacua ${\bf C^*}/{\bf Z}_2$ is
$u \equiv \langle{\rm tr}\phi^2\rangle$.

In this situation the $SU(2)_G$ gauge symmetry is
broken down to an Abelian subgroup $U(1)$ and
the global ${\bf Z}_8$ symmetry is broken to ${\bf Z}_4$.
The spontaneously broken ${\bf Z}_2$ acts on
$u$-plane by $u\to -u$.

\subsection{Wick rotation to Euclidean theory}

In order to carry out instanton calculations
it is necessary that the theory is Wick-rotated to
a Euclidean space.
The Wick rotation is defined in momentum space so that
integrations over the time component $p^0$ of a momentum
changes by rotating +90 degrees around the origin in
$p^0$ complex plane.
We can make it by $p^0\to ip_4$, or equivalently by $t\to-it$.
The gauge field $A_m$ is also Wick-rotated.
The time component is Wick-rotated to
 $A_0({\bf x},t)\to iA_0({\bf x},-it) \equiv iA_4({\bf x},t)$
so as to form a consistent covariant derivative
$\partial_0 - iA_0 \to i(\partial_4 - iA_4)$.
In addition, the invariant tensors
$\sigma^m{}_{\alpha\dot\beta}$ and
$\overline\sigma^{m\dot\alpha\beta}$
are conveniently redefined.
Among other prescriptions, we take
$\sigma_m{}_{\alpha\dot\beta}=(-1,-i\tau^a)$
$\overline\sigma_m{}^{\dot\alpha\beta}=(-1,i\tau^a)$,
from which we have $\overline\lambda\overline\sigma^mD_m\lambda
\to i\overline\lambda\overline\sigma_mD_m\lambda$.
We note that the relation
$\overline\sigma_m{}^{\dot\alpha\beta} =
\sigma_m{}^{\beta\dot\alpha}$
holds, but
$\overline\sigma_m{}^{\dot\alpha\beta}
\ne (\sigma_m{}^{\alpha\dot\beta})^*$.
Auxiliary fields are also Wick-rotated by $D\to iD$, and so on,
so as to produce $\delta$-functional.
After all, the resultant quantities always have lower indices,
if any.

The Wess-Bagger notations after the Wick-rotation will be found
in the Appendices A and B of \cite{Wess-Bagger} by letting
$\eta_{mn}\to-\delta_{mn}$ and
$\epsilon^{mnkl}\to-i\epsilon_{mnkl}$ with $\epsilon_{1234}=+1$;
for example,
$\eta_{mn}\sigma^m{}_{\alpha\dot\alpha}
\overline\sigma^{n\dot\beta\beta} \to
- \sigma_{m\alpha\dot\alpha}\overline\sigma_m^{\dot\beta\beta}=
-2\:\delta_\alpha^\beta\:\delta_{\dot\alpha}^{\dot\beta}$.

Then, the Wick-Rotated Lagrangian is
\begin{eqnarray} \label{LE}
{\cal L}_E &\equiv& - {\cal L} \nonumber\\
&=&
\frac{1}{4}F_{mn}F_{mn}
+ D_m\overline\phi D_m\phi +
\frac{1}{2}[\phi,\overline\phi]^2 \nonumber\\
&&
- \overline\lambda_{\dot\alpha i}
   \overline\sigma_m^{\dot\alpha\alpha}
   D_m \lambda_\alpha{}^i
- \frac{i}{\sqrt{2}}\overline\phi
   \epsilon_{ij}[\lambda^i,\lambda^j]
+ \frac{i}{\sqrt{2}}\phi
   \epsilon^{ij}[\overline\lambda_i,\overline\lambda_j]
{}~,
\end{eqnarray}
where $\lambda_\alpha{}^i = (\lambda_\alpha,\psi_\alpha)$.
Now, let us find the ${\cal N}=2$ supersymmetry transformations.
The ${\cal N}=2$ transformations consist of
a $SU(2)_I$ invariant combination of
two ${\cal N}=1$ supersymmetry transformations
generated by $Q_\alpha^1$ and $Q_\alpha^2$.
In terms of the ${\cal N}=1$ language,
the supercharge $Q_\alpha^1$ acts on
$\lambda_\alpha^1$ and $\lambda_\alpha^2$
as gaugino and Higgsino, respectively,
while the supercharge $Q_\alpha^2$ acts on them
as Higgsino and gaugino, respectively. (cf. (\ref{diamond}))
Then, the supersymmetry transformations are given,
splitting them into holomorphic and anti-holomorphic parts
$\delta = \delta_\xi + \delta_{\overline\xi}$, by
\begin{equation}
\begin{array}{lcllcl}
\delta_\xi A_m &=&
   \overline\lambda_{\dot\alpha i}
   \overline\sigma_m^{\dot\alpha\alpha}\xi_\alpha{}^i ~,~~~
&
\delta_{\overline\xi} A_m &=&
   - \overline\xi_{\dot\alpha i}
   \overline\sigma_m^{\dot\alpha\alpha}\lambda_\alpha{}^i ~,~~~\\
\delta_\xi\lambda_\alpha{}^i &=&
   - \sigma_{mn}{}_\alpha{}^\beta
   \xi_\beta{}^i F_{mn} - \xi_\alpha{}^i D ~,~~~
&
\delta_{\overline\xi} \lambda_\alpha{}^i &=&
   \sqrt{2}\epsilon^{ij}\sigma_{m\alpha\dot\beta}
   \overline\xi^{\dot\beta}{}_j D_m\phi ~,~~~\\
\delta_\xi \overline\lambda_{\dot\alpha i} &=&
   \sqrt{2}\epsilon_{ij}\overline\sigma_{m\dot\alpha\beta}
   \xi^{\beta j} D_m\overline\phi ~,~~~
&
\delta_{\overline\xi}\overline\lambda_{\dot\alpha i} &=&
   \overline\sigma_{mn}{}^{\dot\beta}{}_{\dot\alpha}
   \overline\xi_{\dot\beta i} F_{mn} + \overline\xi_{\dot\alpha i} D ~,~~~\\
\delta_\xi \phi &=&
   -\sqrt{2}\epsilon_{ij}\xi^{\alpha i}\lambda_\alpha{}^j ~,~~~
&
\delta_{\overline\xi} \phi &=& 0 ~,~~~\\
\delta_\xi \overline\phi &=& 0 ~,~~~
&
\delta_{\overline\xi} \overline\phi &=&
   -\sqrt{2}\epsilon^{ij}\overline\xi^{\dot\alpha}{}_i
   \overline\lambda_{\dot\alpha j} ~,~~~\\
\end{array}
\end{equation}
where $D = -i[\phi,\overline\phi]$.

\subsection{twisting}

After the Wick-rotation the Lorentz group of the space-time changes
to a compact rotation group $K$ with appropriate actions,
which is locally $SU(2)_L\times SU(2)_R$.
The connected component of the global symmetry group
is $SU(2)_I$ as seen above.
Therefore, the theory has a global symmetry
$H = SU(2)_L \times SU(2)_R \times SU(2)_I$.
The supercharges $Q_\alpha{}^i$ and $\overline Q_{\dot\alpha j}$
transform under $H$ as
$(2,1,2)$ and $(1,2,2)$, respectively.

Instead of the standard embedding of $K$ in $H$,
we can find two alternative embeddings.\cite{Witten}
As for self-dual instanton calculations the following embedding
is the natural one.
Let $SU(2)_{R'}$ be a diagonal subgroup of
$SU(2)_R \times SU(2)_I$ obtained by sending
$SU(2)_I$ index $i$ to dotted index $\dot\alpha$.
We declare the new rotation group be
$K' = SU(2)_L \times SU(2)_{R'}$.
The supercharges $Q_\alpha{}^i$ and $\overline Q_{\dot\alpha j}$
transform under $K'$ as
$(2,2)$ and $(1,1) \oplus (1,3)$, respectively.
The $K'$ scalar component of the supercharges is interpreted as
a BRST charge $Q_B$.
An interesting point is that, even fermions,
$SU(2)_I$ doublet fields have integer spin with respect to $K'$,
breaking the spin-statistics balance.
Explicitly, we decompose the gaugino doublet $\lambda_\alpha{}^i$
into $K'$ irreducible representations as
\begin{eqnarray} \label{twisted fermion}
\lambda_{\alpha\dot\beta}
&=&
\frac{1}{\sqrt{2}}\sigma_{m\alpha\dot\beta}\:\psi_m
{}~,\nonumber\\
\overline\lambda_{\dot\alpha\dot\beta}
&=&
\frac{1}{\sqrt{2}}\Big(
   \overline\sigma_{mn\dot\alpha\dot\beta}\:\chi_{mn} +
   \epsilon_{\dot\alpha\dot\beta}\:\eta
\Big)~.
\end{eqnarray}
Substituting Eq.~(\ref{twisted fermion})
into the Lagrangian (\ref{LE}), we have
\begin{eqnarray} \label{twisted L}
{\cal L}_E &=& {\cal L}_1 + {\cal L}_2 + {\cal L}_3 \nonumber\\
{\cal L}_1 &=&
\frac{1}{4}F_{mn}F_{mn}
- \chi_{mn}(D_m\psi_n - D_n\psi_m )
+ \frac{i}{\sqrt{2}}\phi\{\chi_{mn},\chi_{mn}\}
{}~,\nonumber\\
{\cal L}_2 &=&
- \eta D_m\psi_m + D_m\overline\phi D_m\phi
- \frac{i}{\sqrt{2}}\overline\phi\{\psi_m,\psi_m\}
{}~,\nonumber\\
{\cal L}_3 &=&
\frac{1}{2}[\phi,\overline\phi]^2  +
\frac{i}{\sqrt{2}}\phi\{\eta,\eta\}
{}~.
\end{eqnarray}

The topological BRST transformation $\delta_B$ induced by
the ${\cal N}=2$ supertransformations is found by putting
$\xi=0$,
$\overline\xi=\epsilon_{\dot\alpha\dot\beta}\rho/\sqrt{2}$,
which reads
$\delta = \delta_\xi + \delta_{\overline\xi} = -\rho \delta_B$.
We have the actions of the BRST transformation $\delta_B$ take
simple forms to satisfy a nilpotency condition
$\delta_B^2=0$ on gauge invariant operators.
We introduce a multiplier of anti-self-dual field $H_{mn}$
with an appropriate BRST transformation laws.
In this case the kinetic term of the Yang-Mills gauge field
becomes
$
\frac{1}{4}F_{mn}F_{mn} =
\frac{1}{2}H_{mn}H_{mn} + iH_{mn}F_{mn}^- +
\frac{1}{4}F_{mn}\tilde{F}_{mn}$.
Here, $F_{mn}^- \equiv \frac{1}{2}(F_{mn}-\tilde{F}_{mn})$.
Now, the BRST transformations are
\begin{eqnarray} \label{BRST transf}
\begin{array}{lcl}
\delta_B A_m &=& \psi_m ~,\\
\delta_B \psi_m &=& -\sqrt{2}D_m\phi ~,\\
\delta_B \chi_{mn} &=& i H_{mn} ~,\\
\delta_B H_{mn} &=& \sqrt{2}[\chi_{mn},\psi] ~,\\
\delta_B \eta &=& i[\phi,\overline\phi] ~,\\
\delta_B \phi &=& 0 ~,\\
\delta_B \overline\phi &=& -\sqrt{2}\eta ~.
\end{array}
\end{eqnarray}
These BRST transformations are nilpotent
up to a gauge transformation
$\delta_B^2 = -\sqrt{2}\delta_\phi$;
e.g., $\delta_B^2A_m = -\sqrt{2}D_m\phi$.
With these BRST transformations at hand,
the ${\cal N}=2$ supersymmetric Yang-Mills Lagrangian
becomes BRST exact up to a surface term:\cite{Witten}
\begin{eqnarray}
{\cal L}_1 &=&
-i\delta_B~\chi_{mn}\Big(iF_{mn}^- + \frac{1}{2}H_{mn}\Big)
+\frac{1}{4}F_{mn}\tilde{F}_{mn}
{}~,\nonumber\\
{\cal L}_2 &=&
\frac{1}{\sqrt{2}}\delta_B~ \overline\phi D_m \psi_m
{}~,\nonumber\\
{\cal L}_3 &=&
-\frac{i}{2}\delta_B~ \eta [\phi,\overline\phi] ~.
\end{eqnarray}

These BRST exact forms are supported from a manifestation of
the ${\cal N}=2$ supersymmetry.
The ${\cal N}=2$ supersymmetric Yang-Mills Lagrangian
is written as an integration $\int d^4 \theta$ over
four chiral superfields $\theta_\alpha^i$.
One combination of the superfields $\theta_\alpha^i$ gives
the topological BRST transformation
$\int d\theta = \partial/\partial\theta$.
Then, the integration over this combination of the superfields
ensures that the Lagrangian takes a BRST exact form.

\section{Quantization}
\label{Q}

In this section we will demonstrate path integrations
in the twisted topological theory.
For a class of Green functions miraculous cancellations occur
between bosonic and fermionic excitations.
And ultimately, an infinite dimensional path integration
reduces to a finite dimensional integration over
a supersymmetric instanton moduli space.

\subsection{instanton moduli space}

An instanton moduli space ${\cal M}_k$ for
a given instanton number $k$
is defined by the self-dual equation $F_{mn}=\tilde{F}_{mn}$
modulo small gauge transformation
which can be continuously deformed to the identity transformation
at infinity.
For instanton calculations, however, it is convenient to include
degrees of freedom corresponding to global gauge transformations.
So, the moduli space ${\cal M}_k$ has dimension $8k$
for a generic $SU(2)$ instanton.

Given an instanton solution $A_m$, we can find a solution
$A_m+\delta A_m$ of its infinitesimal deformation by solving
\begin{equation} \label{self-dual}
\delta F_{mn}^- = (D_m\delta A_n - D_n\delta A_m)^- = 0 ~.
\end{equation}
For an anti-symmetric tensor $X_{mn}$ we define
$X_{mn}^- \equiv \frac{1}{2}(X_{mn}-\tilde{X}_{mn})$.
In addition, we are interested in solutions $\delta A_m$
orthogonal to the directions of gauge orbit.
We impose the gauge fixing condition
\begin{equation} \label{inst GF}
D_m \delta A_m = 0 ~,
\end{equation}
where the covariant derivative is defined
in terms of the instanton background $A_m$.
The condition (\ref{inst GF}) does not exclude modes of
``global'' gauge transformations $\delta A_m=D_m\theta$
satisfying $D_mD_m\theta=0$.
The number of solutions of
Eqs.~(\ref{self-dual}) and (\ref{inst GF}) is
the dimension of the moduli space $dim({\cal M}_k)$.

The equations of motion for the field $\psi_m$ are given
by differentiating the Lagrangian (\ref{twisted L})
with respect to $\chi_{mn}$ and $\eta$.
If we restrict field configurations to instanton solutions
and omitting higher order terms in $g^2$, we have
\begin{eqnarray} \label{psi eqs}
(D_m\psi_n-D_n\psi_m)^- &=& 0 ~,\nonumber\\
D_m\psi_m &=& 0 ~.
\end{eqnarray}
As will be seen in the subsection~\ref{VS},
the fermionic zero modes are enough to be evaluated
in the weak gauge coupling limit $g\to0$.
Comparing Eqs.~(\ref{psi eqs}) with
Eqs.~(\ref{self-dual}) and (\ref{inst GF}),
we have an interesting relation that
fermionic zero modes $\psi_m$ are one-form
on the instanton moduli spaces; $\psi_m\sim\delta A_m$.
We already encounter this type of relation
in Eq.~(\ref{BRST transf}).
Namely, the BRST transformation $\delta_BA_m=\psi_m$ gives
a tangent vector of the moduli space.

An index theorem tells us that the number of $\psi_m$ zero modes
minus the number of $\chi_{mn}$ and $\eta$ zero modes is equal to
the dimension of the moduli space $dim({\cal M}_k)$.
In the present case of a generic $SU(2)$ instanton,
there are no $\chi_{mn}$ and $\eta$ zero modes.
Then, the number of $\psi_m$ zero modes is equal to
the dimension of the moduli space.

The Lagrangian has a $U(1)_R$ symmetry which counts
the number of gauginos and Higgsinos;
$\psi$ has charge $U=1$, and $\chi_{mn}$ and $\eta$ have $U=-1$.
Therefore, anomaly of $U(1)_R$ charge $U$ gives
$\psi_m$ fermionic zero modes.

If we neglect the effect of gauge symmetry breaking,
the dimension of the instanton moduli space is $8k$ and
there are $8k$ fermionic $\psi_m$ zero modes.
In the Coulomb branch, we have a gauge symmetry breaking
caused by the Higgs vacuum expectation value (\ref{vev}).
In this branch the dimension of bosonic true zero modes is $4$
and there are $4$ fermionic $\psi_m$ zero modes.

\subsection{norm cancellations}\label{nc}

Dadda-DiVechia\cite{DaDi} used a background Feynman gauge
but usually a background Landau gauge\cite{'t Hooft} is used
for instanton, as well as super-instanton, calculations.
The calculations are not carried out
in a supersymmetric manner,
since the gauge fixing procedure violates supersymmetry.
It is not a trivial fact that in a different gauge
the one-loop determinants of bosons and fermions
cancel out completely.

For our convenience,
in an instanton background a Landau gauge fixing condition
\begin{equation} \label{GF}
D_m A_m = 0
\end{equation}
is imposed besides the Wess-Zumino gauge.
This condition (\ref{GF}) implies (\ref{inst GF}).
We should hold Eqs.~(\ref{inst GF}) and (\ref{psi eqs})
to keep a useful relation that
$\psi_m$ is a tangent vector of the instanton moduli space.
Then, we add a BRST exact gauge fixing term
\begin{equation} \label{L4}
{\cal L}_4 = -iB D_m A_m
+ \overline c D_m (\partial_m c -i[A_m,c])
\end{equation}
to the original Lagrangian (\ref{twisted L}).
Notice that the BRST charge in this case is stemmed from
the usual gauge transformation, and is different from
the topological one (\ref{BRST transf}).

If we use only a single topological BRST transformation
as in \cite{LaPe,BaSi},
we have ${\cal L}_4=-\delta_B\overline c D_mA_m$
up to a null term.
In this case we have $\delta_B^2=0$.
Such nilpotent BRST transformation consist of a linear combination
of (\ref{BRST transf}) and the usually defined one
(e.g. $\delta_B A_m=D_m c$, $\delta_B \psi_m=i\{c,\psi\}$ and so on).
We list first a few:
\begin{eqnarray}
\delta_B A_m    &=& \psi_m + D_m c ~,\nonumber\\
\delta_B \psi_m &=& -\sqrt{2}D_m\phi + i\{c,\psi\} ~,\nonumber\\
\delta_B c      &=& \sqrt{2}\phi + \frac{i}{2}\{c,c\} ~,\nonumber\\
&\cdots& ~.
\end{eqnarray}
Even if we use these BRST transformations, the following argument
does not change since ${\cal L}_4$ takes the same form.

Now, the quadratic part of the Lagrangian is
\begin{eqnarray}
{\cal L}_{\rm quad} &=&
\frac{1}{2}H_{mn}^2
+ iH_{mn}(D_mA_n - D_nA_m - \epsilon_{mnkl}D_kA_l) \nonumber\\
&&
- \chi_{mn}(D_m\psi_n - D_n\psi_m
- \epsilon_{mnkl}D_k\psi_l) \nonumber\\
&&
+ D_m\overline\phi D_m\phi + \overline c D_m D_m c
- \eta D_m\psi_m - iB D_m A_m \nonumber\\
&\equiv& \Phi \Delta_B \Phi - \Psi D_F \Psi ~,
\end{eqnarray}
where $\Phi = (A_m,H_{mn},\phi,\overline\phi,B)$ and
$\Psi = (\psi_m,\chi_{mn},c,\overline c,\eta )$.
Notice that the covariant derivative is evaluated
in a self-dual instanton background.

The operators $D_F$ and $\Delta_B$ have non-zero modes
defined by
$D_F\Psi=\lambda\Psi$ and $\Delta_B\Phi=\lambda^2\Phi$,
which are specified by a non-zero eigenvalue $\lambda\ne0$.
The path integration over such non-zero modes is given by
bosonic and fermionic Gaussian integrations with respect to
coefficients with which path integration variables
$\Phi$ and $\Psi$ are expanded by the eigenstates.

To guarantee the norm cancellation in a definite way,
the space-time should have a finite volume
and the same boundary condition should be imposed on both
the fermion field $\Psi$ and the boson field $\Phi$.
The finite volume space-time implies that the eigenstates of
the operators $\Delta_B$ and $D_F$ are countable.
The same boundary condition leads to the same spectrum for
the two operators.

Under this situation we may expect that
the norm cancellations between
the bosons $\Phi$ and the fermions $\Psi$
for non-zero modes are guaranteed by
the quartet mechanism.\cite{KO,KU}
Namely, if we calculate
the Green function $\langle{\cal O}\rangle$
of a physical observable satisfying $\delta_B {\cal O}=0$,
path integration over such quartet states
does not contribute to the Green function.
Therefore, we can integrate out any BRST quartet modes
by simply throwing away such modes.

Before proceeding to an explanation of norm cancellations,
let us show the three quartet multiplets.
We split $A_m$ into two parts.
The gauge fixing condition (\ref{GF}) defines
a transverse part $A_m^{^{(T)}}$.
Fluctuations along gauge orbit defines a longitudinal part
$A_m^{^{(L)}}$, and can be written
in terms of a $SU(2)$ gauge group valued function $\varphi(x)$
by $A_m^{^{(L)}}=D_m\varphi$.
These two parts are orthogonal to each other:
\[
\int d^4x A_m^{^{(L)}}A_m^{^{(T)}}
= \int d^4x D_m\varphi\cdot A_m^{^{(T)}}
= -\int d^4x\varphi D_mA_m^{^{(T)}}
= 0 ~.
\]
The topological BRST symmetry leads to
one quartet multiplet
$(A_m^{^{(T)}},\psi_m^{^{(T)}};\chi_{mn},H_{mn})$.
The usual gauge symmetry defines the usual quartet multiplet
$(A_m^{^{(L)}},c;\overline c,B)$,
as well as $(\psi_m^{^{(L)}},\phi;\overline\phi,\eta)$.
In the present theory the quartet members are seen schematically
in Table~\ref{tab:quartets}.
\begin{table}[thbp]
\begin{center}
\begin{tabular}{cccccc}
$U =$ & $-2$ & $-1$ & $0$ & $1$ & $2$ \\
\hline
    &    &    & $A_m^{^{(T)}}$  & & \\
    &    &$\chi$ &  & $\psi_m^{^{(T)}}$ & \\
    &    &    & $H$ &   &   \\
\hline
    &    &    &   & $\psi_m^{^{(L)}}$ &   \\
& $\overline\phi$ &    &   &   & $\phi$  \\
    &    & $\eta$ & &   &   \\
\hline
    &    &    & $A_m^{^{(L)}}$ &   &   \\
&&$\overline c$ &   & $c$ &   \\
    &    &    & $B$ &   &   \\
\hline
\end{tabular}
\vspace{0.3cm}
\caption[]{
The three quartets of
the twisted ${\cal N}=2$ supersymmetric Yang-Mills theory.
$U$ is ghost number which is identical to
the $U(1)_R$ charge.
}\label{tab:quartets}
\end{center}
\end{table}

In the case of calculating the partition function,
the Gaussian integrations give
\begin{equation} \label{Sdet=1}
\frac{{\rm Pf}'(D_F/g^2)}
{\sqrt{{\rm Det}{'(\Delta_B/g^2)}}} = 1 ~.
\end{equation}
This can be shown explicitly as follows.
First of all, coupling constants appearing
in the fermion and boson determinants cancel,
since we have the same number of
the massive degrees of freedom
for the fermions and the bosons.
So, we can omit the coupling constant
in the partition function by setting $g^2=1$.
Secondly, the integrations over the usual ghosts
$c$ and $\overline c$ give ${\rm Det}'(D_mD_m)$,
while the $\phi$ and $\overline\phi$ integrations give
its inverse $1/{\rm Det}'(D_mD_m)$.
So, these modes are canceled.
Thirdly, the integrations over the multipliers
$B$ and $\eta$ simply give $\delta$-functionals
with gauge fixing conditions $D_mA_m=0$ and $D_m\psi_m=0$,
respectively.
The modes $A_m^{^{(L)}}$ and $\psi_m^{^{(L)}}$ integrate
such $\delta$-functionals canceling the results of
themselves.
Finally, the remaining Lagrangian can be rewritten
in terms of a matrix notation
$\frac{1}{2}\Phi^TM_B\Phi-\Psi^TM_F\Psi$ where
$\Phi=(A_l^{^{(T)}},H_{mn})$ and
$\Psi=(\psi_l^{^{(T)}},\chi_{mn})$, and
\begin{equation}
M_B =
\left(\begin{array}{c|c}
0 & i\stackrel{\leftarrow}X \\
\hline
iX & 1
\end{array}\right)
 ~, \qquad
M_F =
\left(\begin{array}{c|c}
0 & -\stackrel{\leftarrow}X \\
\hline
X & 0
\end{array}\right) ~,
\end{equation}
with $X \equiv \delta_{ln}D_m-\delta_{lm}D_n+\epsilon_{lmnk}D_k$.
These matrices have a transpose symmetry property
$M_B^T=M_B$ and $M_F^T=-M_F$.
Because of gauge fixing conditions,
the fields $A_m^{^{(T)}}$ and $\psi_m^{^{(T)}}$ have
the same degrees of freedom as $H_{mn}$ and $\chi_{mn}$,
respectively.
So, the matrix $X$ has a determinant.
Thus, these forms of the matrices allow us to find the relation
\begin{equation}
\sqrt{{\rm Det'}M_B} = {\rm Pf}'M_F = {\rm Det'}X ~.
\end{equation}
Therefore, we have Eq.~(\ref{Sdet=1}).

This is a special case of norm cancellations between
the bosonic and fermionic excitations by supersymmetry
in microscopic one-loop calculatoins.\cite{DaDi}
If we work with no instanton background,
the relation (\ref{Sdet=1}) holds
for all the small fluctuations.

\subsection{partition function and observable}

For calculation of any Green functions the result
does not change by twistings,
since twistings are merely a change of
path integration variables, and we do not modify
the metric of the original flat space-time.
The twisting approach becomes very efficient when we calculate
a Green function of operators that are lowest component of
a ${\cal N}=2$ chiral multiplet.
In terms of the twisted topological field theory,
such operators are BRST singlet.
We call this type of Green function chiral Green function.
If we calculate a chiral Green function,
the right twisting given by (\ref{twisted fermion})
is the appropriate choice.
In this case all good properties
of the twisted topological field theory can be inherited.

Let us recall properties of the twisted topological field theory.
The Lagrangian of
the ${\cal N}=2$ supersymmetric Yang-Mills theory
consists of two terms.
The first one takes a BRST exact form $\{Q_B,V\}$, and
the other one is a surface term
$\frac{1}{4}F_{mn}\tilde{F}_{mn}$.
Notice that
since the gauge fixing term (\ref{L4}) is also BRST exact,
adding this does not change the following property.
By omitting the surface term, the partition function
\begin{equation}
Z_{_{\rm TFT}} = \int DA ~ e^{-\frac{1}{g^2}\{Q_B,V\}}
\end{equation}
is of the twisted topological field theory.

The physical interpretation of states changes significantly
by twisting.
In terms of the topological BRST operator
the subsidiary condition $Q_B|{\rm phys}\rangle=0$ picks up
the lowest parts of the supermultiplets
in the original supersymmetric theory.
Then, the physical states of the twisted topological theory
will be much smaller than
that of the original supersymmetric theory.

Green functions of the topological field theory does not depend on
the bare coupling constant if the inserted operator is BRST singlet
$\delta_B {\cal O} = 0$.
Namely,
\begin{eqnarray} \label{mvls}
\frac{d}{d\beta}\langle {\cal O} \rangle_{_{\rm TFT}} &=&
\frac{d}{d\beta} \int DA ~ e^{-\beta\{Q_B,V\}} {\cal O}
\nonumber\\ &=&
- \int DA ~ e^{-\beta\{Q_B,V\}} \{Q_B,V{\cal O}\}
\nonumber\\ &=& 0 ~,
\end{eqnarray}
where $\beta = 1/g^2$.
Then, we can evaluate the Green functions
in the week coupling limit $g^2\to 0$.

If observables are singlet with respect to
the BRST charge of the right (or left) twisted theory,
only the self-dual (or anti-self-dual) instantons contribute.
This is seen from the following Eq.~(\ref{sdsp}).

\subsection{vanishing theorem}\label{VS}

The supersymmetric Yang-Mills theory
and its twisted topological theory have
the same perturbative corrections around
a given background field, since the surface term
(in our case, $\frac{1}{4g^2}F_{mn}\tilde{F}_{mn}$)
of the Lagrangian does not affect small fluctuations.
Suppose that we integrate out massive modes leaving
instanton zero modes as a background.
The Wilsonian effective Lagrangian is given in terms of
a coupling expansion series
\begin{equation} \label{Wilsonian}
\frac{1}{g^2}{\cal L}_{\rm eff} =
\frac{1}{g^2}{\cal L}_{\rm tree} + {\cal L}_{\rm 1-loop} +
g^2{\cal L}_{\rm 2-loop} + O(g^4) ~.
\end{equation}

If supersymmetry is preserved quite precisely,
we might expect that the perturbation series terminates
at one-loop.
However, the gauge fixing condition (\ref{inst GF})
even violates supersymmetry
in our case as well as many cases of
(super-) instanton calculations.
So it is not trivial that only up to one-loop perturbation
corrections survive in such super-instanton calculations.

Fortunately, more than one-loop order terms should vanish
in our case,
since we can take the weak coupling limit $g^2\to 0$
by virtue of the topological BRST symmetry.
Then, from (\ref{Wilsonian})
only the tree and one-loop terms are survived.
This reminds us with a result by \cite{HSW} that
in a manifestly supersymmetric calculation the ${\cal N}=2$
Yang-Mills theory has only one-loop divergences.

When we consider a Green function of a BRST singlet observable,
it can be seen that the quantum corrections are exhausted by
saddle points of instantons and
the one-loop fluctuations around them.
Mathai-Quillen formula\cite{MaQu,VaWi} shows that
small fluctuations cancel around a saddle point up to a sign.
A leading term of the topological Lagrangian reads
\begin{equation} \label{sdsp}
{\cal L}_{_{\rm TFT}} =
\{Q_B,V\} =
\frac{1}{2}(F_{mn}^-)^2 + \cdots ~.
\end{equation}
Since the first term is positive definite,
only the self-dual configurations $F_{mn}^-=0$ contribute
to the Green function in the weak coupling limit.
The other terms form BRST quartets
to cancel the small fluctuations around
the instanton saddle points.
This is shown explicitly in subsection~\ref{nc}.
Other contributions than self-dual instantons are suppressed
by $\exp(\frac{-1}{2g^2}(F_{mn}^-)^2+\cdots)$.

Once we know that only the fluctuations around
self-dual (or anti-self-dual) instantons contribute to
a Green function $\langle{\cal O}\rangle_{_{\rm SYM}}$,
we can calculate it using
the right (or left) twisted topological field theory.
We divide the integration region of the gauge field into
subspaces specified by a given instanton number.
In each subspace the surface term
$\frac{1}{4}F_{mn}\tilde{F}_{mn}$ becomes a constant,
and the path integration may be carried out
using the twisted topological Lagrangian.
Then, neglecting the surface term such a calculation gives
a result $\langle{\cal O}\rangle_{_{\rm TFT}}^k$
specified by a given instanton number $k$.
As a result, we have
\begin{equation}
\langle{\cal O}\rangle_{_{\rm SYM}} =
\sum_{k=0}^\infty~ \langle{\cal O}\rangle_{_{\rm TFT}}^k~
e^{-\frac{8\pi^2}{g^2}k} ~.
\end{equation}

We notice that a holomorphy assumption
is not introduced in showing the above fact.
It is non-trivial if observables do not receive
perturbative corrections beyond one-loop
in an instanton background,
which is now shown.\cite{Witten,VaWi}

\section{Supersymmetric Instanton Calculations}
\label{SIC}

We work in a Coulomb branch where gauge symmetry breaking
occurs.
In this type of theories instantons do not appear
in a simple way.
Some zero modes due to instantons are raised
by the Higgs mechanism.\cite{'t Hooft,Affleck,ADS}
We demonstrate a one-instanton calculation in terms of
the twisted topological field theory.

\subsection{BPST instanton in singular gauge}

Let us consider the $k=1$ instanton moduli space.
The collective coordinates are four translations $y_m$
and one dilatation $\rho$, and
it is convenient to include degrees of freedom
which correspond to
residual three global gauge rotations $\theta^a$.
Now, the $k=1$ instanton has eight degrees of freedom.

The BPST instanton\cite{BPST,VZNS} in singular gauge is
\begin{equation}
A_m^a =
2\rho^2 \frac{\overline\eta_{mn}^a x_n}{x^2(x^2+\rho^2)} ~.
\end{equation}
where we suppress translations and gauge rotation
collective coordinates.
The corresponding field strength is
\begin{equation}
F_{mn}^a =
- 4\rho^2 \frac{\overline\eta_{mn}^a}{(x^2+\rho^2)^2}
+ 8\rho^2
\frac{(\overline\eta_{mk}^ax_n-\overline\eta_{nk}^ax_m)x_k}
{x^2(x^2+\rho^2)^2} ~.
\end{equation}

Let us find small fluctuations of collective coordinates
satisfying the gauge fixing condition (\ref{inst GF}).
A naive variation $\delta A_m/\delta\gamma$ with $\gamma$
being a collective coordinate does not satisfy
the gauge fixing condition, but it can satisfy
the condition by making use of
a gauge transformation.\cite{Bernard}
The proper variations take a good form
\begin{eqnarray} \label{Am cc}
\delta_{_T} A_m^a &=&
   F_{mn}^a \delta y_n ~,\nonumber\\
\delta_{_D} A_m^a &=&
   F_{mn}^a \frac{x_n\delta\rho}{\rho} ~,\nonumber\\
\delta_{_G} A_m^a &=&
   F_{mn}^a \Big(-\frac{1}{2}\overline\eta_{nk}^b
                   x_k\delta\theta^b \Big) ~,
\end{eqnarray}
where $\delta_{_T}$, $\delta_{_D}$ and $\delta_{_G}$ mean
the variations to the translations ($y_m\to y_m+\delta y_m$),
dilatation ($\rho\to\rho+\delta\rho$) and
global gauge rotations
($\theta^a\to\theta^a+\delta\theta^a$), respectively.

{}From these expressions we can easily show
that they satisfy the gauge fixing condition
using the equation of motion $D_mF_{mn}=0$ and the
self-duality of the curvature.

\subsection{fermionic zero modes}

To find fermionic zero modes a method is available
in $k=1$ instanton case.
The sweeping-out technique\cite{NSVZ} is that
supersymmetry and superconformal transformations yield
all the fermionic zero modes.
Beyond the $k=1$ case a method of tensor products is used.

In our case fermionic zero modes are defined
by Eqs.~(\ref{psi eqs}).
Let us recall a nice relation $\delta_{_B}A_m=\psi_m$
interpreted that fermionic zero modes form a complete set
of tangent vectors on the instanton moduli space.
Then, we easily obtain them by replacing variations
of the gauge field to Grassmann variables;
$\delta y_m\to\xi_m$, $\delta\rho\to\zeta$ and
$\delta\theta^a\to\zeta^a$.
Thus, from (\ref{Am cc}) we immediately have
\begin{equation}
\psi_m^a = F_{mn}^a\Big(
\xi_n + \frac{x_n}{\rho}\zeta
- \frac{1}{2}\overline\eta_{nk}^bx_k\zeta^b
\Big) ~.
\end{equation}

We notice that the ${\cal N}=2$ supersymmetry
admits to find all of the fermionic zero modes
in generic $k$ instantons.
Letting $\xi^{(r)}$ be $8k$ fermionic collective coordinates,
the solution $\psi_m = \sum_r Z_{rm}\xi^{(r)}$
can be found using a quantity $Z_{rm}$ defined
in \cite{Osborn}.
($Z_{rm}$ satisfies (\ref{psi eqs}) and
$\partial A_m/\partial\gamma_r = Z_{rm} + D_m\Lambda$
with $\gamma_r$ being a bosonic collective coordinate.)

\subsection{Jacobian and measure of the collective coordinates}

After integrating out massive modes, integrations over
zero modes remain.
In an usual formula zero modes are written in terms of
collective coordinates.
It is very convenient to change variables from zero modes to
such collective coordinates in carrying out integrations,
which requires Jacobian
from zero modes to collective coordinates.

The metric of the Yang-Mills functional space is defined by
\begin{equation} \label{YM metric}
||\delta A_m^a ||^2 \equiv
\frac{1}{2\pi}\int d^4x~ \delta A_m^a(x)\delta A_m^a(x) ~.
\end{equation}
This metric gives the $A_m$ path-integration a normalization
\begin{equation}
\int DA~ e^{-\frac{1}{2}\int d^4x ~ A_m^a(x)A_m^a(x)}
= 1 ~.
\end{equation}
Restricting a variation $\delta A_m$ to that of
collective coordinates $\delta_{_T} A_m$, $\delta_{_D} A_m$
and $\delta_{_G} A_m$ in Eq.~(\ref{YM metric}),
the metrics of the bosonic collective coordinates are
obtained as
\begin{eqnarray} \label{BJ}
||\delta_{_T} A_m||^2 &=& 4\pi (\delta y_m)^2 ~,\nonumber\\
||\delta_{_D} A_m||^2 &=& 8\pi \delta\rho^2 ~,\nonumber\\
||\delta_{_G} A_m||^2 &=& 2\pi \rho^2 (\delta\theta^a)^2 ~.
\end{eqnarray}

The integration measures of the collective coordinates
are found using a standard formulation that
a metric $ds^2 = g_{ij}dx_idx_j$
leads to a volume form $\sqrt{{\rm det} g_{ij}}d^nx$.
Then, the path integral measure constrained
on the bosonic collective coordinates are
\begin{equation} \label{bosonic measure}
DA = 2^7\pi^4~ d^3\theta\, d^4y\, d\rho\,\rho^3 ~.
\end{equation}
The $\theta$-integration over the $SU(2)$ group gives
$\int_{SU(2)}d^3\theta = 8\pi^2$.
If we define a unit Haar measure
$d\Omega\equiv d^3\theta/(8\pi^2)$,
the overall constant of
Eq.~(\ref{bosonic measure}) coincides with that of
't~Hooft\cite{'t Hooft} ($2^{10}\pi^6$).

Next, let us derive the measure of the fermionic zero modes.
An easy way to do it is that first we regard
the variables $\xi_m$, $\zeta$ and $\zeta^a$
as commuting variables tentatively,
secondly evaluate the metric of the $\psi_m$ functional
space to obtain a Jacobian $J$ from such commuting zero modes
to collective coordinates,
and then the correct Jacobian of the volume form is $1/J$.
The metric is given by
\begin{eqnarray} \label{FJ}
||\delta\psi_m^a||^2 &\equiv&
\int d^4x~ \delta\psi_m^a(x) \delta\psi_m^a(x) \nonumber\\
&=&
2\pi\bigg[
4\pi(\delta\xi_m)^2 + 8\pi\delta\zeta^2 + 2\pi\rho^2(\delta\zeta^a)^2
\bigg] ~.
\end{eqnarray}
This is an expected result from Eqs.~(\ref{BJ}) by virtue of
the wonderful relation $\delta_B A_m^a = \psi_m^a$.
Then, the integration measure of the fermionic zero modes is
\begin{equation}
D\psi =
\frac{d^4\xi d\zeta d^3\zeta}{(2\pi)^4\times (2^7\pi^4\;\rho^3)} ~.
\end{equation}

Therefore, the desired total measure is
\begin{equation}
DA \times D\psi = \frac{1}{(2\pi)^4}
d^4y d\rho d^3\theta \times
d^4\xi d\zeta d^3\zeta ~.
\end{equation}

It is difficult to find Jacobian factors of
bosonic and fermionic zero modes separately,
and at best those of $k=2$ instanton are found\cite{Osborn}.
Owing to the ${\cal N}=2$ supersymmetry,
at least but this is sufficient,
the total Jacobian of the bosonic and fermionic
zero modes can be found very easily.
The metric of collective coordinate $||\delta\psi_m^a||^2$
is deduced from $||\delta A_m^a||^2$ in the same way
as that for the above $k=1$ case.
For a generic ${\cal N}=2$ supersymmetric instanton
specified by an positive integer $k$
we immediately have a simple result
\begin{equation}\label{Jacobian}
J_{bose} J_{fermi} = \frac{1}{(2\pi)^{4k}} ~.
\end{equation}

\subsection{Higgs field and gauge symmetry breaking}

First, let us consider what is determined from
a leading (tree level) behavior of the Higgs field.
Next, we consider one-loop corrections
to the leading contributions, and that's all.
In our prescription the gauge coupling constant is
a overall factor of the action.
All fields can be regarded as order $O(g)$ quantities.
Thus, the relevant terms of the Lagrangian consist of
up to third order polynomial of fields.

For the Higgs field
the leading equation of motion is $D_mD_m\phi=0$.
This has a solution
\begin{equation} \label{homo-phi}
\phi^a = \frac{x^2}{x^2+\rho^2}\phi_0^a ~,
\end{equation}
where $\phi_0^a$ is a constant.
(This form reminds us with
$\delta_{_G}A_m^a = D_m(\delta\theta^a
x^2/(x^2+\rho^2))$;
the solution of $D_mD_m\phi=0$ can be obtained
from $\delta_{_G}A_m^a$).
Similarly $D_mD_m\overline\phi=0$ also determines
$\overline\phi$ with a constant $\overline\phi_0^a$.
The vacuum expectation values (see (\ref{vev}))
of the Higgs fields $\phi$ and $\overline\phi$
determine their behaviors at infinity,
which allow us to find
$\phi_0^b=(\overline\phi_0^b)^*=(0,0,a)$.
These homogeneous solutions parameterize
the $D$-flat directions of the theory.

An self-dual instanton solution and the super-partner
specified by a positive integer $k$
have $8k$ degrees of freedom, respectively.
Accompanied by the gauge symmetry breaking (\ref{vev}),
$8k-4$ of such zero modes are raised,
which we call quasi-zero modes.
The four true zero modes are overall translations.
An identity
$-\delta y_n\partial_nA_m^a = (F_{mn}^a-D_mA_n^a)\delta y_n$
implies that
the infinitesimal deformation $F_{mn}^a\delta y_n$
is always the true zero mode
satisfying Eqs.~(\ref{self-dual}) and (\ref{inst GF}),
and similarly for the four fermionic zero modes.
The massive modes are
dilatation $\delta\rho$, gauge rotations $\delta\theta$
and its superpartners $\delta\zeta$, $\delta\zeta^a$.
The Higgs kinetic term produces a mass term
of the dilatation mode;
\begin{eqnarray} \label{rho-mass}
S_{\rm Higgs} &=&
\int d^4x ~ D_m\overline\phi{}^a D_m\phi^a \nonumber\\
&=&
\int d^4x \partial_m\Big(\overline\phi{}^a D_m\phi^a\Big)
\nonumber\\
&=&
4\pi^2\overline\phi{}_0^a\phi_0^a\rho^2 ~.
\end{eqnarray}
In addition, Yukawa interaction gives
mass term of the fermionic $\zeta$ and $\zeta^a$ modes;
\begin{eqnarray} \label{zeta-mass}
S_{\rm Yukawa} &=&
   -\frac{i}{\sqrt{2}} \int d^4x ~
   \overline\phi\{\psi_m,\psi_m\} \nonumber\\
&=&
-2\sqrt{2}\pi^2\overline\phi{}_0^a\Big(
   2\rho \zeta\zeta^a -
   \frac{1}{2}\rho^2\epsilon^{abc}\zeta^b\zeta^c
\Big) ~.
\end{eqnarray}
These show that translation modes $y_m$ and
super-translation modes $\xi_m$ remain massless.

Finally, let us consider next order effects in perturbation.
The rest of the perturbative corrections does not exist
as explained before in subsection~\ref{VS}.
The relevant terms of the Higgs' equation of motion is
\begin{equation} \label{Higgs eq}
D_mD_m\phi + \frac{i}{\sqrt{2}}\{\psi_m,\psi_m\}= 0 ~.
\end{equation}
The homogeneous solution of this equation is already given.
To obtain the inhomogeneous solution all massive modes
may be integrated out by simply setting them zero.
Here, the fermionic zero modes are now
$\psi_m^a = F_{mn}^a\xi_n$.
Then, the inhomogeneous solution is\cite{FuTr}
\begin{equation} \label{Hinh}
\phi^a_{\rm inh} = \frac{1}{2\sqrt{2}}\xi_m\psi^a_m ~.
\end{equation}
We can easily check this solution by plugging it
into Eq.~(\ref{Higgs eq}).

\subsection{a $k=1$ instanton calculation}

Famously, supersymmetry sometimes provides us
with remarkable properties.\cite{AKMRV}
One of them is a topological property, that
Green functions becomes space-time independent
if they are of lowest components of chiral superfields.
This fact enables us to calculate them
in the weak coupling limit.
(Needless to say, for such a Green function
of gauge invariant operators factorizes.)

Usually, we impose a gauge fixing condition
which violates supersymmetry.
The standard choice (\ref{GF}) is indeed the case.
Unfortunately, gauge dependence can be involved
in such a Green function
if some of the concerned chiral superfields
are not gauge invariant operators.
No reliable calculation cannot be done
for gauge non-invariant quantities.\cite{NSVZ2}
This fact shows that a operator, say,
$\langle {\rm tr}\phi^2 \rangle$ cannot be considered as
a limit $x\to y$ of
$\langle{\rm tr}\phi(x)\phi(y)\rangle$.
In conclusion, we have to calculate in a gauge invariant form,
whereas, we loose to use the topological property\cite{AKMRV}.

Now, here comes the point.
As for gauge invariant operators calculations
in the weak coupling limit is indeed guaranteed
by the quite remarkable property (\ref{mvls}).
Thus, let us calculate the observable
$\langle {\rm tr}\phi^2 \rangle$.
The path integral formula of the observable
$\langle {\rm tr}\phi^2 \rangle$ is
\begin{equation}
\langle{\rm tr}\phi^2\rangle = \int DA~
e^{-\frac{1}{g^2}{\cal L}_E}~ {\rm tr}\phi^2 ~.
\end{equation}
Here we concentrate on a path integration over
the quantum fluctuations continuously deformed
to a $k=1$ super-instanton configuration.
All the massive modes are gauge-fixed,
apart from instanton quasi-zero modes,
by the self-dual condition $F_{mn}=\tilde{F}_{mn}$.
Then, according to the vanishing theorem,
the path integrations over such modes can be carried out
by simply substituting fields with their corresponding
zero modes or quasi-zero modes.
Therefore, the path integration reduces to
a finite dimensional integration
over an supersymmetric instanton moduli space.\cite{Witten}

In actual calculations the determinants of bosons and fermions
are needed to be regularized.
According to \cite{'t Hooft},
we introduce Pauli-Villars regulators with a mass $\mu$
for each field.
After renormalizations the ultraviolet divergences
are absorbed in the bare coupling $g(\mu)$.
Physical quantities do not depend
on the cutoff parameter $\mu$,
so this parameter should appear
in a renormalization invariant form.
In $k=1$ instanton background of
the ${\cal N}=2$ supersymmetric pure Yang-Mills theory,
such form is precisely $\mu^4\exp(-8\pi^2/g^2)\equiv\Lambda^4$.
Then, the super-determinant in a Pauli-Villars regularization
together with the instanton action is
\begin{equation}
\frac{{\rm Pf'}(D_F)}{\sqrt{{\rm Det'}(\Delta_B)}}
e^{-\frac{8\pi^2}{g^2}} = \Lambda^4 ~.
\end{equation}

Let us compile the results obtained in section~\ref{SIC}
to calculate the $k=1$ super-instanton correction
to the quantity $\langle {\rm tr}\phi^2 \rangle$.
First, the inserted operator ${\rm tr}\phi^2$ saturates
the true fermionic zero modes $\xi_m$
through the inhomogeneous part of the Higgs field (\ref{Hinh}).
Only this inhomogeneous part contribute
to the $\int d^4\xi$ integration.
With the $\int d^4 y$ integration we have
\begin{equation}
\int d^4 y\int d^4\xi~ {\rm tr}\phi_{\rm inh}^2
=
\frac{1}{8} \int d^4yF_{mn}^aF_{mn}^a = 4\pi^2 ~.
\end{equation}
The integrations of fermionic quasi-zero modes lead to
\begin{equation}
\int d\zeta d^3\zeta ~ e^{-\frac{1}{g^2}S_{\rm Yukawa}} =
   \frac{(2\pi)^4}{g^4}
   \overline\phi{}_0^a\overline\phi{}_0^a ~ \rho^3 ~.
\end{equation}
The factor $\rho^3$ usually appears
in an integration measure\cite{'t Hooft,Bernard}
of bosonic collective coordinates.
This difference is stemmed from the definition
of fermionic superconformal modes;
$\zeta^a\leftrightarrow\zeta^a/\rho$.
The $\rho$ integration gives
\begin{equation}
\int d\rho \rho^3 e^{-\frac{1}{g^2}S_{\rm Higgs}} =
\frac{g^4}{2(2\pi)^4\Big(\overline\phi{}_0^a\phi_0^a\Big)^2} ~.
\end{equation}
Finally, an overall factor $\int d^3\theta = 8\pi^2$
is multiplid by.

In conclusion, $k=1$ super-instanton corrections
to the vacuum expectation value
$\langle {\rm tr}\phi^2 \rangle$ is
\begin{equation}
\frac{\Lambda^4}{a^2} ~.
\end{equation}
This implies that instanton corrections prevent
the symmetric phase $a=0$.
Compared with the exact result\cite{SW}
using a expansion series\cite{Matone,KLT,ItYa},
we find\cite{FiPo,DKM,ItSa,FuTr}
\begin{equation} \label{Lambdas}
\Lambda = \frac{1}{2}\Lambda_{\rm SW} ~.
\end{equation}
The quantity $\Lambda$ is
a dimensional transmutation scale determined
by a perturbative one-loop $\beta$ function.
In general, any $\beta$ function of coupling constants
takes the same form up to two-loop order,
as far as mass independent renormalization scheme is used.
Then, the definition of $\Lambda$ does not depend on
a wide class of renormalization schemes.
As for the quantity $\Lambda_{\rm SW}$, this is defined
as a scale of $u\equiv\langle{\rm tr}\phi^2\rangle$ in which
monopoles and dyons become massless.
In other words, $\Lambda_{\rm SW}$ is a zeros
of a low energy effective parameter ${\rm Im}(\tau_{\rm eff})$
along real values of $u$.
So, this quantity is not accessible
by any instanton calculations of finite number $k$.
Then, the relation (\ref{Lambdas}) should be regarded
as an input in instanton calculations.
In order to compare the result of instanton calculations
with a exact result, at least two results
with different instanton number $k$ are required.

In \cite{FiPo,FuTr,DKM} $k=2$ instanton calculations
are demonstrated for
the ${\cal N}=2$ supersymmetric Yang-Mills theory.
Consistency is found between the exact result
and instanton calculations.

\section{Summary and Discussion}

A viewpoint of twisted topological theory\cite{Witten}
is introduced for supersymmetric instanton calculations.
This viewpoint enables us to observe that
non-perturbative contributions
are indeed saturated only by instantons.
Physical observables are defined in terms of a BRST charge
stemmed from ${\cal N}=2$ supersymmetry,
besides they should be gauge invariant.
It is shown that Green functions of physical observables
can be calculated in the weak coupling limit.
As an example we demonstrate such instanton calculation
for a quantity $\langle {\rm tr}\phi^2 \rangle$.

The BRST exact form for the Lagrangian is supported
from a manifestation of the ${\cal N}=2$ supersymmetry.
The ${\cal N}=2$ supersymmetric Yang-Mills Lagrangian
is written as an integration $\int d^4 \theta$ over
four chiral superfields $\theta_\alpha^i$.
One combination of the superfields $\theta_\alpha^i$ gives
the topological BRST transformation
$\int d\theta = \partial/\partial\theta$.
Then, the integration over this combination of the superfields
ensures that the Lagrangian takes a BRST exact form.

We concentrate on a path integration over
the quantum fluctuations continuously deformed
to a supersymmetric instanton configuration.
All the massive modes are found to be gauge-fixed,
apart from instanton quasi-zero modes,
by the self-dual condition $F_{mn}=\tilde{F}_{mn}$.
Then, according to the vanishing theorem
explained in subsection~\ref{VS},
the path integrations over such modes can be carried out
by simply substituting fields with their corresponding
zero modes or quasi-zero modes.
Therefore, the path integration reduces to
a finite dimensional integration
over an supersymmetric instanton moduli space.

In actual calculations gauge fixing is needed.
Dadda-DiVechia\cite{DaDi} showed norm cancellations
in a background Feynman gauge.
While, we impose a background Landau gauge (\ref{GF}),
and this gauge is usually used
for supersymmetric instanton calculations.
If a gauge fixing procedure violates supersymmetry,
norm cancellations are not guaranteed
by supersymmetry completely.
Then, it is necessary to check that in a different gauge
the one-loop determinants of bosons and fermions
cancel out completely.
We show this in subsection~\ref{nc}.

We show that the ${\cal N}=2$ supersymmetry
admits to find all of the fermionic zero modes
in generic $k$ instanton background.

It is difficult to find Jacobian factors of
bosonic and fermionic zero modes separately,
and at best those of $k=2$ instanton are found\cite{Osborn}.
We show that owing to the ${\cal N}=2$ supersymmetry,
the total Jacobian of the bosonic and fermionic
zero modes is found in Eq.~(\ref{Jacobian}).

\clearpage

\begin{center}
\Large Acknowledgements
\end{center}
We would like to thank H. Hata and K. Higashijima
for discussions and suggestions,
T. Harano and M. Sato for helpful comments,
and also thank to Y. Imamura for conversations,
and especially thank to S. Tanimura
for helpful explanations of the Mathai-Quillen formula.
The work was supported in part by
the Research Fellowships of
the Japan Society for the Promotion of Science
for Young Scientists.


\end{document}